\documentclass[floats,floatfix,showpacs,amssymb,prd,twocolumn,superscriptaddress,nofootinbib]{revtex4-1}

\usepackage{amsmath,color,graphicx}
\newcommand\unit[1]{{\rm #1}}

\newcommand\mysec[1]{\vspace{2mm}\noindent \emph{#1}--}

\usepackage{hyperref}
\definecolor{linkcolor}{rgb}{0.0,0.3,0.5}
\definecolor{urlcolor}{rgb}{0.27,0.55,0.}
\definecolor{funcolor}{rgb}{0.65, 0.16, 0.16}
\hypersetup{colorlinks=true,linkcolor=linkcolor,citecolor=linkcolor,filecolor=linkcolor,urlcolor=linkcolor}

\def\apj{\rm{ApJ}}
\def\apjl{\rm{ApJ}}

\def\apjs{\rm{ApJS}}
\def\mnras{\rm{MNRAS}}
\def\prd{\rm{PRD}}
\def\prl{\rm{PRL}}
\def\prx{\rm{PRX}}
\def\cqg{\rm{CQG}}

\def\nat{\rm{Nature}}
\def\aap{\rm{A\&A}}
\graphicspath {{figures/}}

\newcommand\NoSignificance[1]{}
\begin{document}

\title{Inferences about supernova physics from gravitational-wave measurements: GW151226 spin misalignment as an indicator of strong black-hole natal kicks}

\author{Richard O'Shaughnessy}
\email{rossma@rit.edu}
\affiliation{Center for Computational Relativity and Gravitation,
Rochester Institute of Technology, 85 Lomb Memorial Drive, Rochester, NY 14623, USA}
\author{Davide Gerosa}
\thanks{Einstein Fellow}
\email{dgerosa@caltech.edu}
\affiliation{TAPIR 350-17, California Institute of Technology, 1200 E California
Boulevard, Pasadena, CA 91125, USA}
\author{Daniel Wysocki}
\email{dw2081@rit.edu}
\affiliation{Center for Computational Relativity and Gravitation,
Rochester Institute of Technology, 85 Lomb Memorial Drive, Rochester, NY 14623, USA}

\begin{abstract}
The inferred parameters of the binary black hole GW151226 are consistent with nonzero spin for the most massive black
hole,  misaligned from the binary's orbital angular momentum.  If the black holes formed through isolated binary evolution
from an initially aligned binary star,
this misalignment would then arise from a 
natal kick imparted to the first-born black hole at its birth during stellar collapse. 
We use simple kinematic arguments to constrain the characteristic magnitude of this kick, and find that a natal kick $v_k \gtrsim 50$  km/s 
must be imparted to the
black hole at birth
to produce misalignments consistent with GW151226. 
{Such large natal kicks exceed those adopted by default in  most of the current supernova and binary evolution models.}

\end{abstract}
\maketitle
\mysec{Introduction}
The Laser Interferometer
Gravitational Wave Observatory (LIGO)  has reported the  discovery of two  binary black holes (BHs): GW150914 and GW151226
\cite{2016PhRvX...6d1015A}.  
The masses and inferred birth rate of these events are  consistent with prior predictions 
\cite{2016Natur.534..512B,2010ApJ...715L.138B,2012ApJ...759...52D,2016MNRAS.458.2634M,2016A&A...588A..50M},
derived by  assuming these objects form from the evolution of isolated pairs of stars; see, e.g.,
\cite{2016ApJ...818L..22A}. 
At this early stage, observations cannot firmly distinguish between this formation channel and other proposed
alternatives, such as the formation of binary BHs in densely interacting clusters  \cite{2016ApJ...824L...8R},
{in gas disks  \cite{2017arXiv170207818M}}, %
or
as primordial BHs \cite{2016PhRvL.116t1301B}. 
If, however, binary BHs do form from isolated binary evolution, then precise measurements of their properties provide unique clues into how BHs and massive stars evolve.  

Assuming BH binaries form
from  initially aligned binary stars (i.e., all angular momenta are parallel),  
the most likely processes that can misalign their spin
angular momenta 
are the linear momentum recoils imparted when a BH's progenitor star
ends its life in a supernova (SN)  \cite{2013PhRvD..87j4028G,2000ApJ...541..319K}.  
Observations strongly suggest that asymmetries in the SN process can indeed impart \NoSignificance{significant} strong natal kicks to newly formed compact objects.  Based on the proper motion measurements of pulsars in the Milky Way, it is believed that supernovae (SNe) can impart velocities as high as $v_k\sim 
450$ km/s   to neutron stars \cite{2005MNRAS.360..974H}.  
Conversely, the occurrence of natal kicks onto BHs is less clear. On the one hand, observations of galactic $x$-ray binaries suggest that
 BH natal kicks may be as large as hundreds of km/s
\cite{2012MNRAS.425.2799R,2015MNRAS.453.3341R,2017MNRAS.467..298R,2016ApJ...819..108B,2010MNRAS.401.1514M,2012ApJ...747..111W,2014ApJ...790..119W}.  
On the other hand, natal kicks onto heavier BHs could be significantly reduced, as their very massive progenitor stars are expected to undergo prompt
collapse and not eject enough material to enable strong recoils (see, e.g.,
\cite{2012ARNPS..62..407J} and references therein).  
Measurements of natal kicks through electromagnetic observations have already been proved crucial to understand the
physics of SNe. For instance, if BH kicks are indeed as large as those imparted to neutron stars, this would require
large-scale asymmetries of the SN ejecta %
\cite{2013MNRAS.434.1355J,2017ApJ...837...84J},    {or  anisotropic neutrino emission during collapse
  \cite{1998ApJ...495L.103L,2015MNRAS.453.3341R,2006ApJS..163..335F}.} 

Gravitational wave (GW) measurements of merging binary BHs have the potential to provide crucial insights on this issue.
SN kicks can reach (or even exceed) the expected orbital velocities of the stellar binary from which 
binary BHs formed with dramatic effects on its 
formation and evolution.
Strong natal kicks disrupt many potential compact binary
progenitors (thus affecting the expected GW rates \cite{2016Natur.534..512B,1999A&A...346...91B}) and drastically tilt the orbital plane of the
few that survive (which greatly affects the spin precession dynamics by the time the source becomes visible in LIGO \cite{2013PhRvD..87j4028G,2000ApJ...541..319K}).
Several previous studies have demonstrated  that the GW signature of BH spin-orbit
misalignments can be efficiently identified \cite{2014PhRvD..89j2005O,2014PhRvL.112y1101V,2015PhRvD..91d2003V,2016PhRvD..93d4071T} 
and used to distinguish between formation channels \cite{2013PhRvD..87j4028G,2017CQGra..34cLT01V,2017arXiv170306873S}.  
We point out two  examples. %
First, LIGO provides strong constraints on a  quantity that is both nearly conserved on astrophysical time scales \cite{2008PhRvD..78d4021R,2015PhRvL.114h1103K,2015PhRvD..92f4016G} and of key astrophysical interest: the effective spin 
$\chi_{\rm eff}  = \hat{\mathbf{L}} \cdot(\mathbf{S}_1/m_1 + \mathbf{S}_2/m_2)/(m_1+m_2)$, where $m_{1,2}$ and $\mathbf{S}_{1,2}$ are the masses and spins of the component BHs, and $\mathbf{L}$ is the binary's orbital angular momentum
(we used natural units $G=c=1$).
BH binaries assembled in densely interacting environments have random spin orientations and thus $\chi_{\rm
  eff}$ is frequently negative, while binaries formed in isolation from initially aligned stellar progenitors are expected to be found with positive effective spin \cite{2016ApJ...832L...2R}. Second, for binaries formed in isolation, the azimuthal projection of the BH spins onto the orbital plane $\Delta\Phi$  was found to directly track the occurrence of mass transfer and tidal spin alignment between the stellar progenitors \cite{2013PhRvD..87j4028G,2016PhRvD..93d4071T,2014PhRvD..89l4025G}. 
In this Letter, we use simple kinematic arguments to draw conclusions about the strength of SN kicks from the reported
 observation of GW151226.  This is the less massive of the two confirmed GW detections, where nonzero natal kicks are more likely.  Leaving complicated binary evolution physics aside, 
we show how to translate the  spin misalignments reported by LIGO into concrete constraints on the strength of the first
SN kick.  
{This approach already proved successful. Assuming misaligned jets observed in $x$-ray binaries
  \cite{1995Natur.375..464H,1997ApJ...477..876O,2001ApJ...555..489O} are good indicators of the BH spin direction,
  \citet{2010MNRAS.401.1514M} used similar kinematic arguments to constrain the natal kick imparted to the microquasar
  GRO J1655-40. They find the observed misalignment $\gtrsim 10^\circ$ can only be explained with a kick of few tens of
  {km/s}.
}

\mysec{Observations of GW151226}
The LIGO and Virgo Collaborations characterized GW151226  as a binary BH, with component masses
$14.2_{-3.7}^{+8.3} M_\odot$ and $7.5_{-2.3}^{+2.3}M_\odot$ \cite{2016PhRvL.116x1103A}.   The right panel of their Fig. 4 provides a
posterior distribution on the magnitude and orientation of the two BH spins, relative to the orbital angular
momentum.   Their analysis suggests both that  the more massive BH likely had nonzero spin
and, critically,  that this spin was most likely modestly misaligned with the orbital angular momentum, with a
misalignment angle $\gamma$ ranging between $25^\circ$ and $80^\circ$.

Because of significant precession, the spin-orbit misalignments that LIGO directly measures and reports, corresponding to GW frequencies of  $20~\unit{Hz}$, in
  principle must be evolved backwards in time to identify the spin orientations when the BHs  first
  formed \cite{2015PhRvL.114h1103K,2015PhRvD..92f4016G}.  
Although this process turns out to be crucial to  extract
  astrophysical information from full GW data, its details are not important for this study where we only focus on loose
  constraints on the measured spin direction (i.e. $25^\circ\lesssim\gamma\lesssim80^\circ$). 
Moreover, in the simple assumption adopted here where additional alignment processes (such as tidal interactions) are
neglected, previous work showed there is no net tendency to align or antialign the BH spins \cite{2013PhRvD..87j4028G}.
{This point will be specifically addressed in future work.}

\mysec{Formation and misalignment of GW151226 from isolated evolution}
GW151226 could have formed from the evolution of a pair of isolated massive stars ``in the field'' \cite{2016PhRvX...6d1015A}. Concrete
formation scenarios for this event can be easily extracted from exhaustive simulations of binary evolution over cosmic
time \cite{2016Natur.534..512B} [the evolutionary scenarios described here are drawn from the publicly available  ``Synthetic Universe'' (\href{http://www.syntheticuniverse.org/}{www.syntheticuniverse.org})]. 
As a representative example, GW151226 could have formed from a pair of $53M_\odot$ and $25 M_\odot$ stars, initially in
a relatively close and modestly elliptical orbit with semimajor axis $R=4000R_\odot$; as the stars evolve and the more massive star transfers and loses mass, the binary evolves to a
$22 M_\odot$ helium star and a $26 M_\odot$ companion in a modestly tighter and circularized orbit of $900 R_\odot$;  the primary then
undergoes a SN explosion, losing a small amount of mass to form a $19.7 M_\odot$ BH.   The kick following this first explosion tilts the orbital plane, changing
relative alignment between the orbital plane and the BH's spin direction -- presumed to be parallel to the preexplosion orbital angular
momentum. %
 Subsequent phases of stellar interaction, notably, when the BH spirals through the envelope of the secondary
 star, stripping it and leaving behind a helium core {while itself accreting $0.455 M_\odot$}, cause the binary to progress to a much tighter circular orbit of a few $R_\odot$ prior
 to the second SN.   
Because the common-evelope phase typically shrinks the orbital separation of a factor $\gtrsim 100$, the orbital velocity $v=\sqrt{GM/R}$ (where $M$ is the binary's total mass) at the second SN event is typically an order of magnitude
 larger than the velocity prior to the first; {in this case, $R=6.6 R_\odot$ and $v=1090\unit{km/s}$}. 
 Since the effect of the kick onto the binary only depends on the ratio $v_k/v$ (see below), this second SN has a minimal impact on the misalignment of the
 orbital angular momentum \cite{2013PhRvD..87j4028G}.  
If SN kicks are indeed responsible for the observed misaligned primary BH in GW151226, it is likely this formed during the first SN.
Moreover, the first-born BH accretes too little  matter to appreciably change its angular momentum direction, even
during the common-envelope phase
\cite{2007ApJ...662..504B,2008ApJ...682..474B,1999MNRAS.305..654K,1998ApJ...506L..97N}.

\mysec{Spin-orbit misalignment from natal kicks}
The orbital-plane tilt angle introduced by the first SN kick can be calculated using simple Newtonian kinematics
\cite{2013PhRvD..87j4028G,2000ApJ...541..319K}.  For simplicity, here we only study the typical case in which strong
binary interactions {like tides and mass transfer} have circularized the orbit {and aligned the stellar spins before the first SN} \cite{1981A&A....99..126H}.  We likewise assume for simplicity that
the initially most massive object undergoes the first SN explosion, {and that the SN explosion itself does not
  torque the  BH}.  If $\mathbf{r}=
\mathbf{r}_2 - \mathbf{r}_1$ is the relative orbital separation, $\mathbf{v}=d\mathbf{r}/dt$ is the orbital
velocity and $\mathbf{v}_k$ is the imparted kick velocity, then the orbital angular momentum per unit reduced mass changes from $\mathbf{L}/\mu=\mathbf{r} \times\mathbf{v}$ to 
$\mathbf{L}_{f}/\mu_f=\mathbf{r} \times (\mathbf{v} +  \mathbf{v}_k)$, where $\mu_f\neq \mu$ because of mass loss during the explosion.
The orbital plane tilt $\gamma$  reads
\begin{align}
\cos \gamma = {\mathbf{\hat L}} \cdot {\mathbf{\hat L}_f} =
\frac{(\mathbf{v} + \mathbf{v}_k)\cdot \mathbf{\hat v}}{\sqrt{(\mathbf{v} + \mathbf{v}_k\cdot \mathbf{\hat v})^2 + (\mathbf{v}_k\cdot\mathbf{\hat L})^2 }}\,.
\end{align}
Assuming the spin of the collapsing star $\mathbf{S}$ was aligned to the orbital angular momentum before the explosion (i.e. $\mathbf{\hat S} = \mathbf{\hat L}$), $\gamma$ also equals the spin misalignment angle of the newly formed BH.  If the kick imparted by the explosion is sufficiently large, the post-SN eccentricity exceeds unity and the binary does not remain bound.
 If $\beta = M_f/M$ denotes the fraction of total mass retained by the binary after the 
explosion, disruption occurs if $F(\mathbf{v}_k)<0$ where 
\begin{eqnarray}
F(\mathbf{v}_k) = 2 \beta - 1 - \frac{|\mathbf{v}_k|^2}{v^2} - 2 \frac{\mathbf{v}_k\cdot \mathbf{v}}{v^2}\,.
\end{eqnarray}
Finally,  the cumulative distribution
of the misalignment angle $\gamma$ between pre and post-SN angular momenta can be expressed as 
\begin{eqnarray}
P(\gamma<\gamma_*)=\frac{\int d\mathbf{v}_k \,p(\mathbf{v}_k)  
  \Theta[ \gamma_* - \gamma(\mathbf{v}_k) ]
 \Theta[F(\mathbf{v}_k)]
}{
\int d\mathbf{v}_k\, p(\mathbf{v}_k) \Theta[F(\mathbf{v}_k)]
}
\end{eqnarray}
where $\Theta(x)$ is the Heaviside step function and $p(\mathbf{v}_k)$ is the kick velocity probability distribution. For simplicity, in the following we assume that $p(\mathbf{v}_k)$  is an isotropic Maxwellian distribution
characterized by a single one-dimensional width $\sigma$ (corresponding to a mean square velocity $\langle v_k^2\rangle = 3 \sigma^3$), as found for neutron stars \cite{2005MNRAS.360..974H}.
Motivated by the formation scenario illustrated  above, we assume modest mass
loss in SN explosions,  adopting $\beta=0.98$ as a representative example of the  narrow range of $\beta$ found
in typical population-synthesis studies ($0.95-1$) \cite{2016Natur.534..512B}; we stress that this choice does not significantly influence our results.   %

\begin{figure*}
\centering
\includegraphics[height=0.375\textwidth]{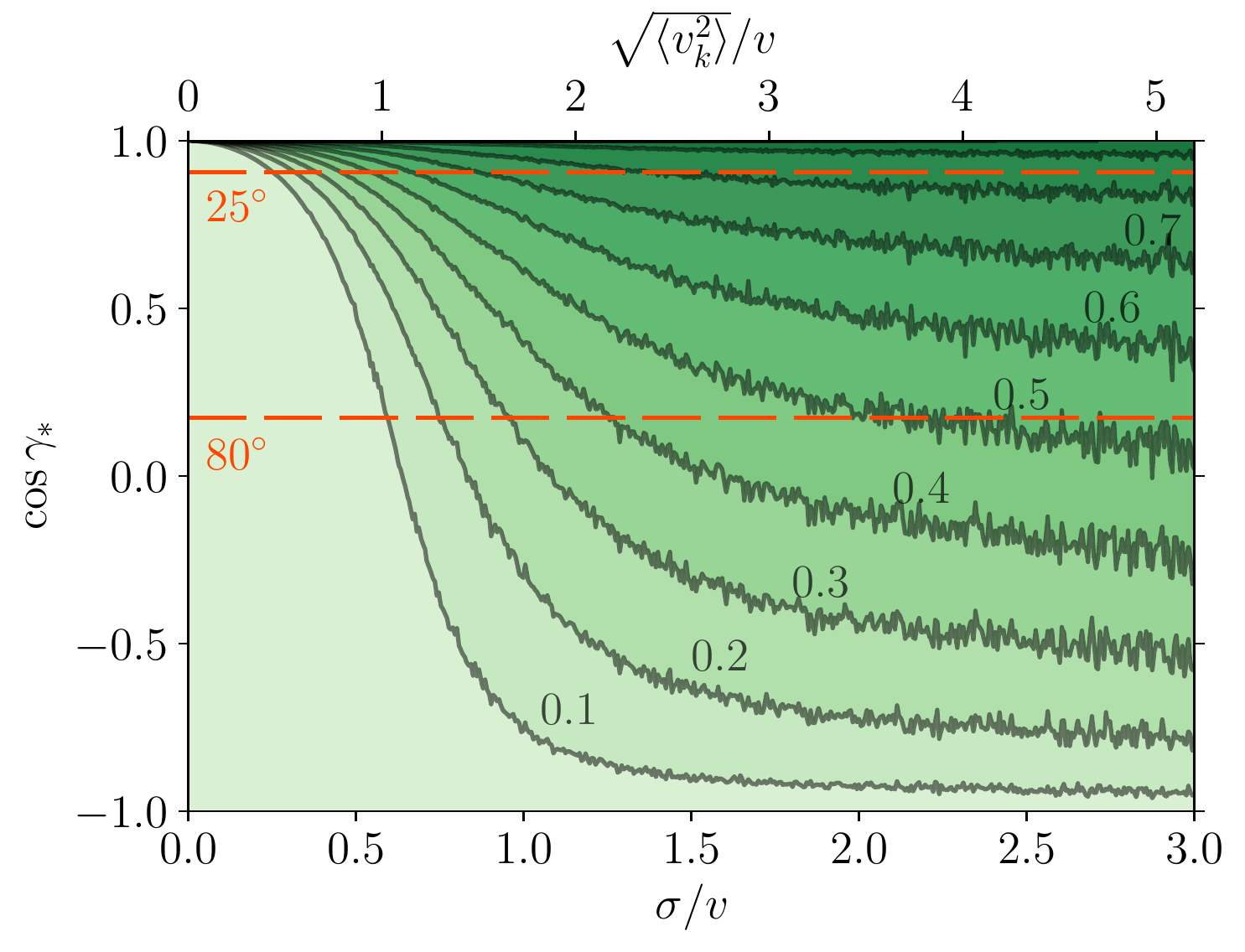}
$\;\;$
\includegraphics[height=0.375\textwidth]{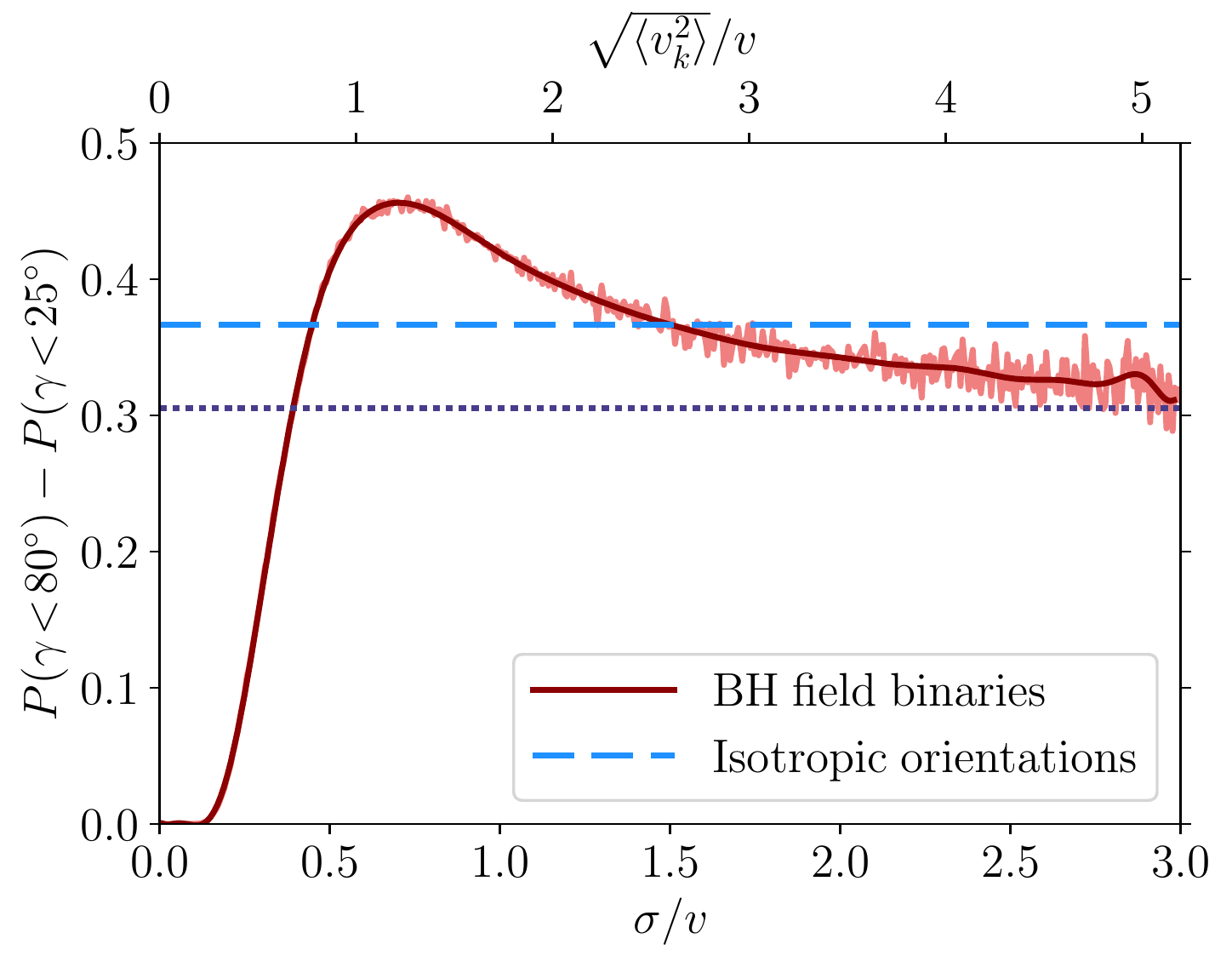}
\caption{\label{fig:TypcialMisalignment}\textbf{Comparing kick-induced misalignments with GW151226.} \emph{Left:} Contour plot of the cumulative
  probability distribution $P(\gamma<\gamma_*)$ of the spin misalignment $\gamma$
  produced by the first SN kick in a binary similar to the progenitor of GW151226. The natal kick is assumed to be drawn from a Maxwellian
  distribution characterized by $\sigma$, which enters our predictions only through its ratio with the binary orbital velocity $v$. For a sense of scale, horizontal dashed lines are drawn at $\gamma=25^\circ$ and
  $\gamma=80^\circ$ as found {as upper and lower bounds for} for GW151226 \cite{2016PhRvX...6d1015A}.  
\emph{Right}: Fraction of surviving binaries  with spin misalignment  consistent with GW151226 as a function of the dimensionless kick magnitude $\sigma/v$. The lighter pink line shows $P(\gamma<80^\circ) - P(\gamma<25^\circ)$ from our Monte Carlo runs, while the darker red curve shows a polynomial fit. For context, the horizontal dashed line
shows $(\cos 25^\circ - \cos 80 ^\circ)/2$, as expected from random spin-orbit {mis}alignment, while the horizontal dotted line
  corresponds to $(80^\circ-25^\circ)/180^\circ$, as expected in the limit of large $\sigma$.
As natal kicks increase in magnitude, the fraction with 
misalignment consistent with GW151226 first increases substantially, as most surviving binaries have been modestly
kicked relative to their orbital speed.
}
\vspace{1.2cm}
\includegraphics[height=0.375\textwidth]{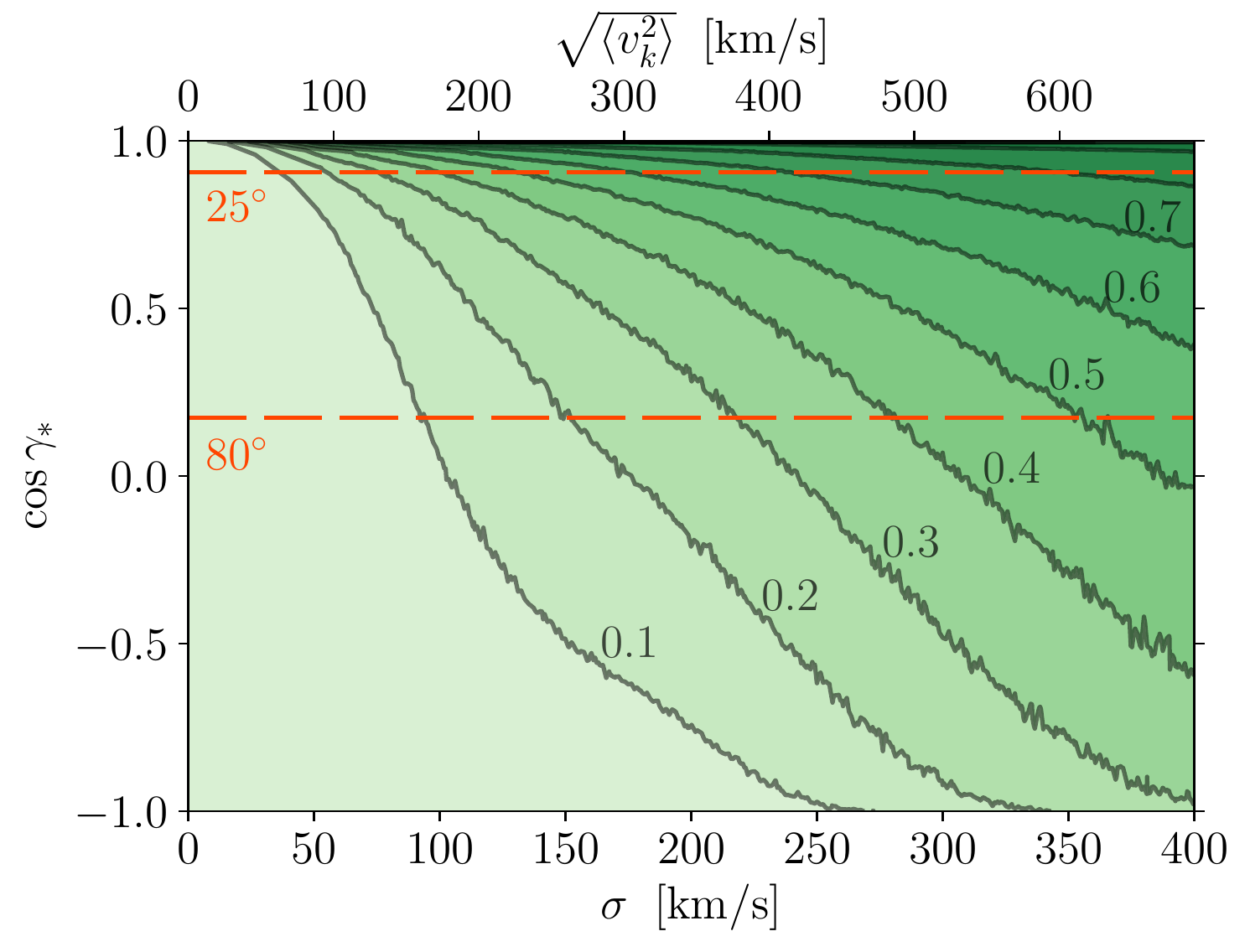}
$\;\;$
\includegraphics[height=0.375\textwidth]{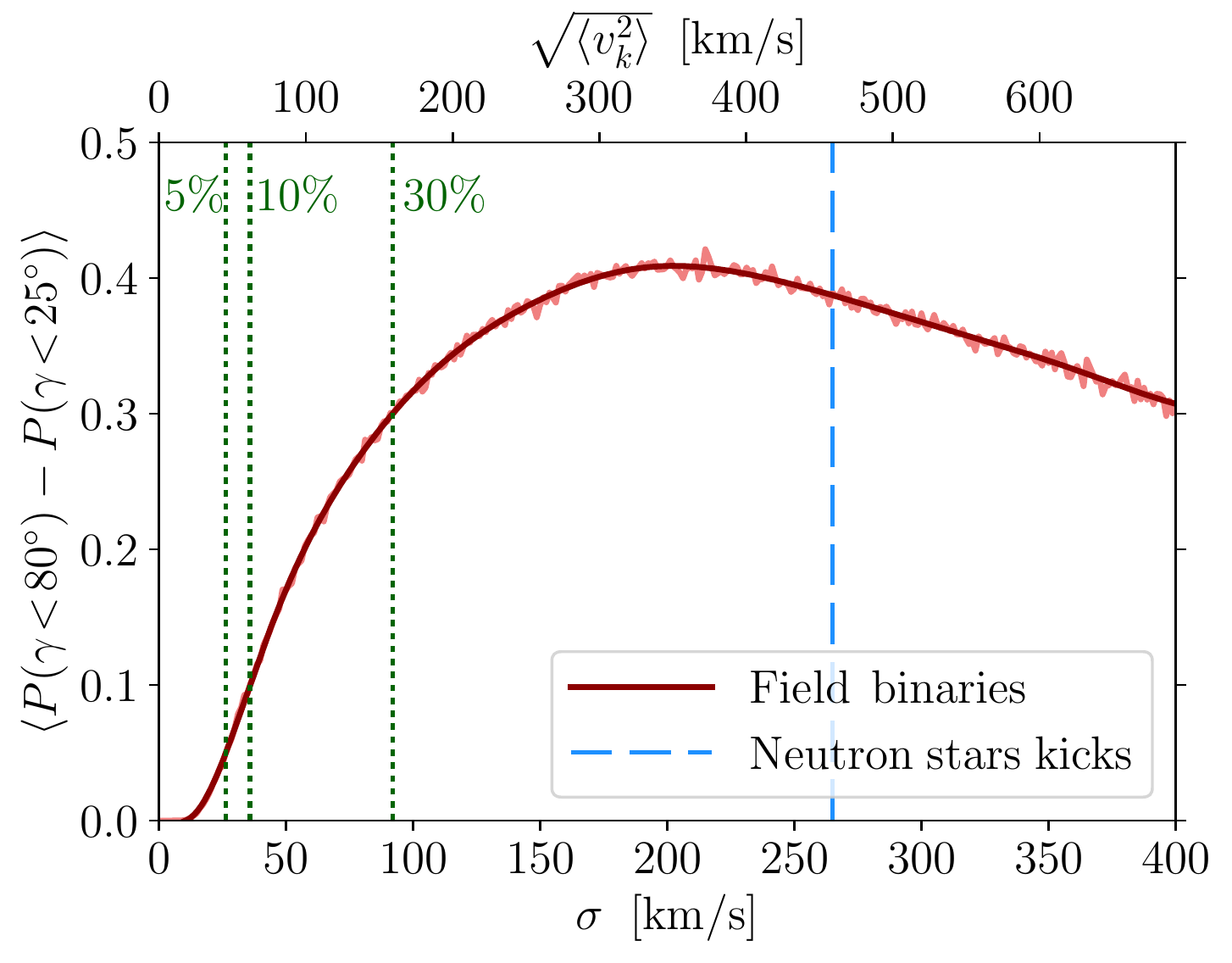}

\caption{\label{fig:Units}\textbf{Kick velocities consistent with GW151226 misalignment}: \emph{Left}: Cumulative
  distribution $\left<P(\gamma<\gamma_*)\right>$ averaged over masses and separations as a function of misalignment
  angle $\cos \gamma$ and  physical natal kick strength $\sigma$. The top axis shows the correspondent three-dimensional root-mean-square velocity $\sqrt{\langle v_k^2\rangle} = \sqrt{3} \sigma$. \emph{Right}: Difference $\left<P(\gamma<80^\circ)-P(\gamma<25^\circ)\right>$ versus $\sigma$ and $\sqrt{\langle v_k^2\rangle}$, illustrating how the expected fraction of
binaries with misalignments consistent with GW15226 changes with the characteristic natal kick magnitude.  The lighter pink line shows results of our Monte Carlo runs, while the darker red curve corresponds to a polynomial fit. Vertical
green dotted lines are drawn at $\sigma\simeq 26, 36$ and 92 km/s, corresponding to probabilities of $5\%, 10\%$ and
$30\%$; the dashed blue line at $\sigma=265$ km/s marks the typical natal kick magnitude imparted to neutron stars
\cite{2005MNRAS.360..974H}.}
\end{figure*}

Because the dimensionless quantities $\gamma$ and $F$ depend on
natal kicks only through the ratio $\mathbf{v}_k/v$, the probability $P(\gamma<\gamma_*)$ depends on $\sigma$ only through the dimensionless ratio $\sigma/v$.
In the limit of large $\sigma/v$, the distribution of misalignments among surviving binaries approaches a nearly uniform
distribution, i.e., $P(\gamma<\gamma_*) \simeq \gamma_*/\pi$. 
The left panel of Fig.~\ref{fig:TypcialMisalignment} shows the misalignment distribution pertinent to GW151226 (i.e., $25^\circ<\gamma<80^\circ$), as a function of the
unknown  dimensionless kick magnitude  $\sigma/v$; for comparison, horizontal lines show the range of misalignments
implied by the LIGO observations. On the right,  we show the probability of a kick misalignment that is both consistent with these
limits \emph{and} does not unbind the orbit. Only modest SN kicks of
$\sigma \gtrsim 0.5 v$ allow a wide range of spin-orbit misalignments consistent with GW151226.

To convert from a relative to an absolute velocity scale, we adopt a distribution of progenitor masses and separations
consistent with GW151226 and with observations of massive stars
\cite{2016Natur.534..512B,2012Sci...337..444S}.  We assume that the binary is
circular; the
primary mass is drawn from a power-law distribution $p(m_1) \propto m_1^{-2.35}$ between $30 M_\odot$ and $100
M_\odot$,  $m_2$ is drawn from a
uniform distribution between $20 M_\odot$ and $m_1$,  and the orbital period $P_{\rm orb}$ is drawn from a distribution $p(P_{\rm orb})\propto (\log
P_{\rm orb}/\unit{day})^{-0.5}$, with limits set by twice the radius of the stars of interest ($R=40 R_\odot$)  %
and by the maximum radius of one of the two stars' giant phase ($R= 3\times 10^3 R_\odot$).
We then compute the ensemble-averaged cumulative probability distribution
\begin{align}
\left<P(\gamma<\gamma_*) \right>  = \int &P( \gamma<\gamma_*| m_1,m_2,P_{\rm orb},\sigma)p(m_1)p(m_2) \nonumber \\
  & \times p(P_{\rm orb})dm_1 dm_2 dP_{\rm orb}
\end{align}
For simplicity, we neglect mass transfer before the first SN and assume that all binaries which survive the first SN kick are equally likely to form a binary BH similar to GW151226. 
To the extent it holds, our calculations can be applied to generic  binary BHs formed from isolated evolution, not just GW151226.

Figure \ref{fig:Units} shows the distribution of kick
misalignments as a function of $\sigma$.  As expected given the characteristic velocity of bound orbits of massive
stars, a natal kick of  at least $\sqrt{\langle v_k^2\rangle}\simeq 45$ (62) km/s must be imparted to the first-born BH
to obtain the misalignment of GW151226 in $5\%$ ($10\%$) of the realizations. If BH natal kicks are as large as those imparted to neutron stars ($\sigma\simeq 265$ km/s \cite{2005MNRAS.360..974H}), up to $\sim 39\%$  of our realizations are found consistent with the observed spin misalignment.

\mysec{Distinguishing from alternative models}
Coalescing binary BHs could form in dense interacting environments, where the spin and orbital angular
momentum directions are randomized, i.e. $P(\gamma<\gamma_*) = \cos\gamma_*/2$.  The right panel of Fig.~\ref{fig:TypcialMisalignment} also shows a horizontal line corresponding to the probability  that a randomly oriented binary will lie within
the region observed for GW151226.
Field binaries with $0.5\lesssim \sigma/v \lesssim 1.5$  have a higher probability to produce misalignment consistent with GW151226 than binaries formed through dynamical interactions. 
 As pointed out in
\cite{2016ApJ...832L...2R}, modest SN kicks cannot produce an isotropic spin distribution. As $\sigma$ increases, the misalignment distribution becomes uniform in $\gamma$, below the randomly oriented result (which predicts a distribution uniform in $\cos\gamma$).
However, strong SN kicks $\sigma \gg 2 v$ on BHs both disrupt most field binaries and eject BHs from globular clusters, dramatically reducing the merger rate and creating difficulties for any stellar-evolution-based formation scenarios. %
While SN kicks can more easily explain the observed spin-orbit misalignments for the particular case of GW151226,
 observations of the spin misalignment \emph{distribution} from
many future events are crucial to support or rule out different formation scenarios.

The Kullback-Leibler (KL) divergence $D_{KL} = \int d\gamma  p(\gamma) \ln[ p(\gamma)/q(\gamma) ]$ provides 
a measure of the difference
between two distributions $p(\gamma),q(\gamma)$, and  hence the number of detections needed before we can
distinguish between models (i.e., $N \simeq 1/D_{KL}$) \cite{2013PhRvD..88h4061O}. We can calculate the KL
divergence between the isotropic spin misalignment distribution and the distributions implied by any $\sigma/v$ shown in
Fig.~\ref{fig:TypcialMisalignment} or any $\sigma$ shown in Fig.~\ref{fig:Units}.
 Even loosely accounting for  measurement error (e.g., using the width of the distribution of
GW151226 as an estimate of the relative misalignment accuracy), we find  that $\mathcal{O}(10)$ events similar to GW151226 are needed to distinguish between an
isotropic distribution and a distribution misaligned by natal BH kicks, in agreement with other estimates  \cite{2017arXiv170306873S,2017arXiv170601385F}

\mysec{Discussion}
LIGO should detect several  hundred more binary BHs over the next five years  \cite{2016ApJ...833L...1A,2016PhRvX...6d1015A}.
These observations will support or rule out whether binaries
are born with spin strictly aligned with their orbital angular momentum or obtained significant misalignment from natal kicks. They will also provide strong constraints on the strength of such  kicks.  
Relatively low-mass binaries like GW151226 provide the simplest, cleanest laboratory to study the impact of SN
 kicks.  First and foremost, the  explosions that form them are not expected to result from direct collapse
 \cite{2016Natur.534..512B}, so some residual linear momentum will be imparted to the ejected material and the BHs.  Second, low-mass, unequal-mass-ratio binaries like GW151226  accumulate many precession cycles prior to merger
in LIGO's sensitive band \cite{1994PhRvD..49.6274A}.  Third, this regime of precessing inspiral is relatively
well modeled theoretically \cite{1994PhRvD..49.6274A,2014PhRvD..89d4021L,2014PhRvL.113o1101H,2014PhRvD..89f1502T,2015PhRvL.114h1103K,2015PhRvD..92f4016G}, 
and accessible with current parameter-estimation techniques
\cite{2015PhRvD..91d2003V,2016PhRvD..93d4071T,2017CQGra..34cLT01V}.  LIGO has therefore the best chance to make precise measurements about
misalignment for low-mass binaries, where the merger phase is relatively unimportant.
By contrast, for more  massive BHs like GW150914, fewer cycles are available in the LIGO band and the merger phase becomes crucial 
\cite{2016PhRvD..94f4035A,2016PhRvL.116x1102A}.
Phenomenological models that approximate full solutions of Einstein's equations are known to omit important physics, which can in turn lead to  biases when these models are applied to
parameter estimation \cite{2017CQGra..34j4002A}. Robust spin-orbit misalignment measurements for heavy BHs will require
improved  waveform modeling, more extensive use of numerical relativity data
\cite{2016PhRvD..94f4035A,2015PhRvL.115l1102B}, 
{and incremental improvements in low-frequency GW detector sensitivity}.  
The natal kicks required to explain the misalignment of GW151226 are in excess of the fallback-suppressed kicks adopted by default in current binary evolution models %
\cite{2016Natur.534..512B,2016ApJ...819..108B,2016ApJ...832L...2R} (though note models M4, M5, and M6 in \cite{2016Natur.534..512B}). 
 {Notably, these natal kicks are consistent with observations of recoil velocities 
\cite{2012MNRAS.425.2799R,2015MNRAS.453.3341R,2017MNRAS.467..298R,2016ApJ...819..108B}, {and  jet misalignments
  \cite{1995Natur.375..464H,1997ApJ...477..876O,2001ApJ...555..489O,2010MNRAS.401.1514M} of Galactic $x$-ray binaries.}}

For isolated binary evolution models, a modest increase in SN kicks diminishes the expected event rate -- more binary
BHs are disrupted by the first SN --  but otherwise produces predictions for the population of merging
binary BHs that are consistent with
existing observations \cite{2016Natur.534..512B,2016A&A...594A..97B}.  
The impact of recent physically motivated prescriptions that relate natal kick magnitude and ejected mass \cite{2016MNRAS.461.3747B,2013MNRAS.434.1355J} has yet to be fully explored with large-scale population-synthesis studies.

Large natal kicks $\sqrt{\langle v_k^2\rangle}\gtrsim 50$ km/s  that must be imparted to  BHs of mass $\gtrsim 15 M_\odot$ at formation could be  a significant
challenge for SN physics.  
For example, one of the leading models used to explain the kicks imparted to neutron stars invokes gravitational attraction  by the newly formed compact object of some of the material ejected asymmetrically during the explosion (the so-called ``gravitational
tug-boat mechanism" \cite{2013MNRAS.434.1355J,2017ApJ...837...84J}). While this requires
significant and quite asymmetric mass ejection, many of the formation scenarios explored for GW151226 assume very modest mass  loss ($\beta\sim 0.98$), with most of the material
falling back on to (and slowing down) a protoneutron star core that later collapses to a BH (see, e.g.,
\cite{2012ApJ...757...91B,2012ApJ...749...91F}).   
{On the other hand, neutrino-driven kicks do not require significant mass loss.}%

Our analysis assumes that SN kicks provide the principal mechanism for binary spin-orbit misalignment in field binaries.  Alternatively, binaries
could be born with primordial spin-orbit misalignment \cite{2014ApJ...785...83A},  or gain comparable misalignment early in their life
via either interactions with  a
tertiary companion \cite{2014ApJ...785...83A} 
 or core-envelope interactions \cite{2013ApJS..208....4P}.  If such misalignment can persist or grow during the long lifetime and many
interactions necessary to form a coalescing BH, then LIGO observations might be an indicator of primordial spin
misalignment processes.  
{Large-scale  surveys of binary stars can determine if such spin-orbit misalignments occur frequently.}
{Conversely, substantial accretion onto the BH could also align the BH spin with the orbit after the first SN \cite{1998ApJ...506L..97N}.}
 \vspace{-0.1cm}
\mysec{Acknowledgements}
We thank  Krzysztof Belczynski, Emanuele Berti, Michael Kesden, Will Farr, Daniel Holz, and Gijs Nelemans for carefully
reading our manuscript, and our anonymous referees for their helpful feedback.  
R.O'S. and D.G. gratefully acknowledge the hospitality of the Aspen Center for Physics, supported by  NSF Grant No. PHY-1066293, where this work was initiated.  R.O'S. is supported by NSF Grants No.  AST-1412449, No. PHY-1505629, and No. PHY-1607520. D.G. is supported by NASA through Einstein Postdoctoral Fellowship Grant No. PF6-170152 awarded by the Chandra \emph{X}-ray Center, which is operated by the Smithsonian Astrophysical Observatory for NASA under Contract No.
NAS8-03060.  
\vspace{-0.4cm}
\bibliography{bibads}
\end{document}